\documentclass[aps,prd,superscriptaddress,twocolumn,preprintnumbers,nofootinbib,showpacs,floatfix]{revtex4}
\usepackage{graphics}
\usepackage{graphicx}
\usepackage{amsmath}
\usepackage{amssymb}
\usepackage{bbm}

\newcommand{\beq}{\begin{equation}}
\newcommand{\eeq}{\end{equation}}
\newcommand{\bea}{\begin{eqnarray}}
\newcommand{\eea}{\end{eqnarray}}

\begin{document}

\title{Constraints on UED from $W^\prime$ searches}

\author{Thomas Flacke}
\affiliation{Institut f\"ur Theoretische Physik und Astrophysik, 
Universit\"at  W\"urzburg, D-97074 W\"urzburg, Germany}
\affiliation{Department of Physics, Korea Advanced Institute of Science and Technology,
335 Gwahak-ro, Yuseong-gu, Daejeon 305-701, Korea}

\author{Arjun Menon}
\affiliation{Department of Physics and Astronomy, University of
  Oregon, Eugene, Oregon 97405, USA}

\author{Zack Sullivan}
\affiliation{Department of Physics, Illinois Institute of Technology,
  Chicago, Illinois 60616-3793, USA}

\date{July 16, 2012}

\begin{abstract}
  We obtain contraints on three Universal Extra Dimensional models
  utilizing limits from the CMS Collaboration on $W^\prime$ production
  and decay into a single-top-quark final state.  We find a weak
  constraint on the Minimal Universal Extra Dimensions model due to
  small Kaluza-Klein number violating terms.  In contrast, the
  $W^\prime$ search puts a strong limit on the size of the Dirac mass
  term of the quarks in Split Universal Extra Dimension models.  In
  Non-minimal Universal Extra Dimension models the $W^\prime$ search
  constrains the splitting between the boundary localized
  kinetic terms of the gauge bosons and the quarks.  Each of these
  bounds can be translated into constraints on the mass
  splitting between the Kaluza-Klein excitations of the $SU(2)$ charged
  quarks and the Klauza-Klein excitations of the $W$ boson.
\end{abstract}

\preprint{IIT-CAPP-12-06}


\maketitle

\section{Introduction}
\label{sec:Intro}

In models of Universal Extra Dimensions
(UED)~\cite{Appelquist:2000nn}, all standard model particles are
promoted to higher dimensional fields propagating in a flat extra
dimensions. In this article we focus on five-dimensional UED models
in which the extra dimension is chosen to be the orbifold $S^1/Z_2$ so
as to obtain chiral zero mode fermions. The residual $Z_2$ parity,
called KK-parity, implies that odd parity Kaluza-Klein (KK) particles
can only be pair produced. In addition, it guarantees the stability of
the lightest KK-odd particle which represents a viable dark matter
candidate.

Electroweak precision measurements \cite{EWPTrefs}, in combination
with the LHC Higgs bounds \cite{LHCHiggsrefs} applied to UED
\cite{UEDHiggsrefs} and flavor physics \cite{flavorref} impose a bound
of $R^{-1}\gtrsim 700$~GeV on the compactification scale. Adding a
requirement that the dark matter relic density observed by WMAP
\cite{WMAPref} is consistent with UED points to a compactification
scale of $1.3 \mathrm{\;TeV}\lesssim R^{-1}\lesssim 1.5\mathrm{\;TeV}$
for the most commonly considered dark matter candidate
\cite{UEDDMnewref} --- the first KK excitation of the $U(1)_Y$ gauge
boson $B^{(1)}$ \cite{MUEDref,UEDDMrefs,UEDcoannrefs}. The collider
phenomenology of the 5D UED model has been discussed in Refs.\
\cite{colliderrefs,UEDreviewref}.

In spite of its minimal field content, UED on $S^1/Z_2$
contains a large number of undetermined parameters beyond the
compactification radius $R$. UED is non-renormalizable and therefore
must be considered as an effective field theory. Naive dimensional
analysis \cite{NDAref}, unitarity of KK mode gauge boson scattering
\cite{unitrefs}, and stability of the Higgs potential vacuum
\cite{vacref} imply that the UED cutoff is of $\mathcal{O}(10)$ times
the compactification scale. Unless the UED UV completion is specified,
this cutoff, as well as parameters of higher dimensional operators,
must be considered as free parameters of the model which have to be
constrained by experiment.

The lowest dimensional operators allowed by all symmetries of the
model are additional kinetic terms which are localized at the orbifold
fixed points, so-called boundary localized kinetic terms
(BLKTs).\footnote{Conservation of KK-parity requires all boundary
  localized operators to be included symmetrically on both fixed
  points.} In the Minimal UED model (MUED) \cite{MUEDref}, BLKTs are chosen
to be zero at the cutoff scale $\Lambda$. At lower scales, they are
induced via one-loop corrections. Non-zero BLKTs affect the UED KK
mass spectrum~\cite{BLKTrefs} as well as couplings amongst different
KK mode particles, and therefore have a large impact on UED
phenomenology. BLKTs can change the lightest Kaluza-Klein particle LKP
from the commonly considered $U(1)$ gauge boson $B^{(1)}$ to the
neutral $SU(2)_L$ gauge boson KK mode
$W^{3(1)}$~\cite{Flacke:2008ne}. Also, the mass splittings between the
LKP and other states at the first KK-level are altered, which has a
strong impact on the relic density \cite{UEDcoannrefs}.  Finally, in
the presence of BLKTs, resonant LKP annihilation through second KK
mode excitations is suppressed,\footnote{The masses of particles at
  the $n$-th KK mode are not given by $\sim n/R$, as they are in
  MUED.} while for MUED these processes play an important
role \cite{UEDDMnewref}.

Another possible source of modifications to the KK mode mass spectrum
are fermion bulk mass terms which are introduced in the so-called
split-UED model (sUED) \cite{sUEDref}. Contrary to BLKTs, such terms are not
radiatively induced, as a plain fermion bulk mass term violates
KK parity.  However, they can be introduced as KK-odd mass terms via a
background field.

In both scenarios, non-minimal UED models with boundary localized
kinetic terms (nUED) as well as split-UED, the UED collider
phenomenology is altered. Cascade decays, commonly considered for UED
collider signatures, are altered due to the modified mass spectrum.
An even more striking signature arises from newly induced
couplings between fermion zero modes and even KK-mode gauge
bosons. These couplings lead to $W^\prime$, $Z^\prime$, and
$\gamma^\prime$ signatures in the electroweak sector or colored
resonance signatures in the QCD sector. In MUED
these signatures occur \cite{secKKref}, but the corresponding  couplings are one-loop suppressed. In split-UED
\cite{sUEDcolrefs} and nUED the couplings are already present at
tree-level, and can be large.

In this article, we determine the bounds on the parameter space of
minimal universal extra dimensions,
of non-minimal UED models with boundary localized kinetic terms
as well as of split-UED from the bounds on $W^\prime$ searches
in the single-top-quark decay channel. In section \ref{sec:setup}, we
review the MUED, split-UED, and nUED model and summarize the respective
couplings and KK mass spectra. In section \ref{sec:constraints}, we
use constraints on $W^\prime$ masses and couplings obtained by the CMS
Collaboration \cite{CMSDATA} to derive constraints on
the MUED, split-UED, and nUED parameter space.

\section{Phenomenological Setup}
\label{sec:setup}

At tree-level in universal extra dimensions, the standard model
fermions, gauge bosons, and Higgs fields are promoted to 5D-fields on
$S^1/Z_2$.  The $Z_2$ orbifold condition allows the standard model
particles to be identified with the zero modes of these 5D
fields. Kaluza-Klein parity is the residual symmetry generated by the
breaking of 5D Lorentz invariance due to the boundary conditions. As a
5D theory, UED is non-renormalizable and additional sets of operators
in the bulk and localized at the boundary can significantly modify the
tree-level UED model. In particular, the coupling of second KK mode
gauge bosons like the $W^{(2)}$ to zero mode quarks is no longer
vanishing as it would be if only the UED bulk terms were considered.

\subsection{Minimal Universal Extra Dimensions}

Minimal universal extra dimensions represent the simplest UED setup
in which one-loop corrections are taken into account. The model
has two additional parameters as compared to the standard model: the
compactification scale $R^{-1}$ and the cutoff scale of the theory
$\Lambda$.  At the scale $\Lambda$ all higher dimensional operators
are assumed to be vanishing; however, renormalization group (RG)
evolution generates such higher dimensional local operators at scales below $\Lambda$. The
$W^{(2)}$ mass in MUED follows from~\cite{MUEDref}
\beq
m_{W^{(2)}}^2 = m_2^2 + \delta m_{W^{(2)}}^2 + \bar{\delta} 
m_{W^{(2)}}^2\,,
\eeq
where the bulk induced correction is
\beq
\delta m^2_{W^{(2)}}=-m^2_2\frac{5}{8}\frac{g^2\zeta(3)}{16\pi^4}\,,
\eeq
the boundary induced correction is
\beq
\bar{\delta}m^2_{W^{(2)}}=m^2_2\frac{15}{2}\frac{g^2}{16\pi^2}\ln\left(\frac{\Lambda^2}{\mu^2}\right)\,,
\eeq
$m_2=2/R$, $\zeta$ is the zeta-function, $\Lambda$ is the cutoff scale, and 
$\mu$ is the renormalization scale.
One-loop corrections also lead to couplings between zero mode 
fermions and the $W^{(2)}$ gauge boson of the form~\cite{MUEDref}  
\beq
g_{002}=\frac{g_{000}}{\sqrt{2}}\left[\frac{\bar{\delta}m^2_{W^{(2)}}}{m^2_2}-
2\frac{\bar{\delta}m_{f_2}}{m_2}\right]\,, \label{g002mued}
\eeq
where $g_{000}$ is the zero mode coupling which is identified with the  standard model coupling and 
\bea
\bar{\delta}m_{f_2}&=&m_2\left(3\frac{g_3^2}{16\pi^2}+\frac{27}{16}\frac{g^2}{16\pi^2}+\frac{1}{16}\frac{g^{\prime 2}}{16\pi^2}\right)\ln\left(\frac{\Lambda^2}{\mu^2}\right)\,.
\eea
The coupling in Eq.~(\ref{g002mued}) arises from RG evolution induced mixing between different KK modes of the
same KK parity. An alternative way of understanding these couplings is
that the RG evolution induces boundary localized operators
that modify the equations of motion and boundary conditions for the KK
modes of the fermions and gauge bosons. As the induced BLKTs  for gauge bosons and fermions differ, the wavefunctions of the zero
mode fermions and the $W^{(2)}$ gauge boson are not orthogonal. Hence,
a coupling between $W^{(2)}$ and left-handed zero mode fermions is
induced.\footnote{The presence of these couplings, as well as
  couplings to all higher even-numbered KK modes is a consequence of
  the breaking of 5D translational invariance due to the boundary
  localized terms. Couplings between zero mode fermions and the
  odd-numbered $W$-boson KK modes are not induced, because they are forbidden
  by KK parity.} Since these effects are only induced at the one-loop
level, the couplings between $W^{(2)}$ and left-handed zero mode
fermions are suppressed.

\subsection{Split Universal Extra Dimensions}

Split universal extra dimensions (split-UED) are a UED extension,  initially proposed to explain cosmic ray
observations~\cite{sUEDref}.  In split-UED, a KK parity odd background
field provides an effective 5D Dirac mass term for the 5D fermions of the
form 
\beq \mathcal{L}_{sUED}^{5D} \supset
\mu\theta(y)\bar{\Psi}{\Psi}\,,
\eeq 
where $\mu$ is the induced mass parameter, and $\theta(y)$ denotes the Heaviside step function.

As the gauge bosons are unaffected by this operator, the mass of
$W^{(n)}$ is
\beq
m^2_{W^{(n)}} = m_n^2 + m_W^2\,,
\label{sUEDmW2}
\eeq
where $m_n = n/R$ and $m_W$ is the standard model $W$ boson
mass. The presence of the bulk mass term modifies the profiles of the KK fermions
in the extra dimension, and in particular the fermion zero
mode. Therefore, overlap integrals between zero mode fermions and even
KK modes of the gauge bosons are non-zero, which for the $W^{(2)}$
leads to a coupling \cite{sUEDcolrefs}
\beq
g_{002}=- \sqrt{2} g_{000}\frac{\mu^2 R^2}{\mu^2 R^2+1} \coth \left( \frac{\mu 
\pi R}{2}
\right)\,.
\label{geffsUED}
\eeq
The KK mass spectrum of the fermions is altered as well. For the first KK mode, the mass is given by  \cite{sUEDcolrefs}
\beq
m_{\Psi^{(1)}}=\sqrt{\mu_\Psi^2+R^{-2}}.
\label{sUEDmQ1}
\eeq

\subsection{Non-Minimal Universal Extra Dimensions}
\label{sec:nUED}

In non-minimal extensions of universal extra dimensions, tree-level
boundary localized operators are included into the model. Parameterizing the fundamental domain of the 
$S^1/Z_2$ as  $-\frac{\pi
  R}{2}\leq y \leq \frac{\pi R}{2}$, the electroweak part of the boundary
action of nUED is given by
\begin{align}
S_{BLT}&=\int d^5 x \Bigl[\delta\Bigl(y+\frac{\pi R}{2}\Bigr)+
\delta\Bigl(y-\frac{\pi R}{2}\Bigr) \Bigr] \nonumber\\
\times\Bigl(& - \frac{r_B}{4}B_{\mu\nu}B^{\mu\nu} -
\frac{r_W}{4}W^a_{\mu\nu}W^{a\mu\nu} + r_H (D^\mu H)^\dagger D_\mu H  \nonumber\\
&   +\mu_b^2 H^\dagger H - \lambda_b (H^\dagger H)^2  
+ r_{\Psi_h}\overline{\Psi}_hi\gamma^\mu D_\mu\Psi_h \Bigr) \,,
 \label{FullBLT}
\end{align}
where the $\Psi_h$ denotes $Q_L$, $U_R$, $D_R$, $L_L$, and $E_R$. The
fermion BLKTs $r_{\Psi_h}$ are three $3\times 3$ Hermitian matrices in
flavor space of mass dimension $-1$. Flavor physics dictates them to be proportional to
the unit matrix.\footnote{In Ref~\cite{Csaki:2010az} it has been shown that fermion mass matrices in split-UED induce FCNC's unless the mass matrices are flavor blind, i.e., proportional to the unit matrix in flavor space. The same arguments hold for fermion BLKTs.} 
In principle, boundary Yukawa
couplings could also be present, but as they suffer from the same
flavor problem and do not affect our later analysis, we set them zero in the
above. Our analysis of $W$ KK modes is only affected by the BLKT of
the $SU(2)$ charged quarks, i.e., the parameter $r_{Q}$. The Higgs
BLKT does not have a sizable effect on the $W$ KK mode masses and only
marginally influences the couplings of KK fermions to KK gauge modes.
The $U(1)$ BLKT does not affect the $W$ KK mode masses and couplings
\cite{Flacke:2008ne}. For concreteness, in what follows we set
$r_H=r_B=r_W$, $\mu_b=0=\lambda_b$ and restrict ourselves to positive BLKTs.\footnote{For negative gauge- or fermion BLKTs, the KK spectrum contains unphysical modes (ghosts and/or tachyons).}

Under these assumptions the $W^{(n)}$ mass is 
\beq
m^2_{W^{(n)}}= k_n^2 + m_W^2\,,
\label{Wnmass}
\eeq
where $k_n$ is determined by the quantization condition \cite{Flacke:2008ne},
\bea
r_Wk_n &=& -\tan\left(\frac{k_n \pi R}{2}\right), \mbox{for even $n$ and } \label{k2cond}\\
r_Wk_n &=& \cot\left(\frac{k_n \pi R}{2}\right), \mbox{for odd $n$.} \label{k1cond}
\eea
Using the modified boundary conditions we also find
the coupling of the $W^{(2)}$ KK mode to zero mode quarks to be \cite{FlackeP}
\beq
g_{002}=g_{000}\frac{\sqrt{8}\left(r_W-r_Q\right)}{\pi R +2 r_Q}\sqrt{\frac{1+\frac{2 r_W}{\pi R}}{\sec^2\left(\frac{k_2 \pi R}{2}\right)+\frac{2 r_W}{\pi R}}}\,.
\label{geffnUED}
\eeq 
As can be seen, this KK number violating coupling vanishes for
$r_W=r_Q$.\footnote{In this case, the KK decomposition of the fermion and the gauge fields yields identical wave function bases $\{ f^W_n(y)\}=\{f^Q_n(y)\}$, and the orthogonality relations of the wave functions guarantee the absence of KK number violating operators also for couplings including both $Q$ and $W$ KK modes.}

\section{The CMS $W^\prime$ Constraint}
\label{sec:constraints}

\begin{figure}
\begin{center}
\includegraphics[width=.45\textwidth]{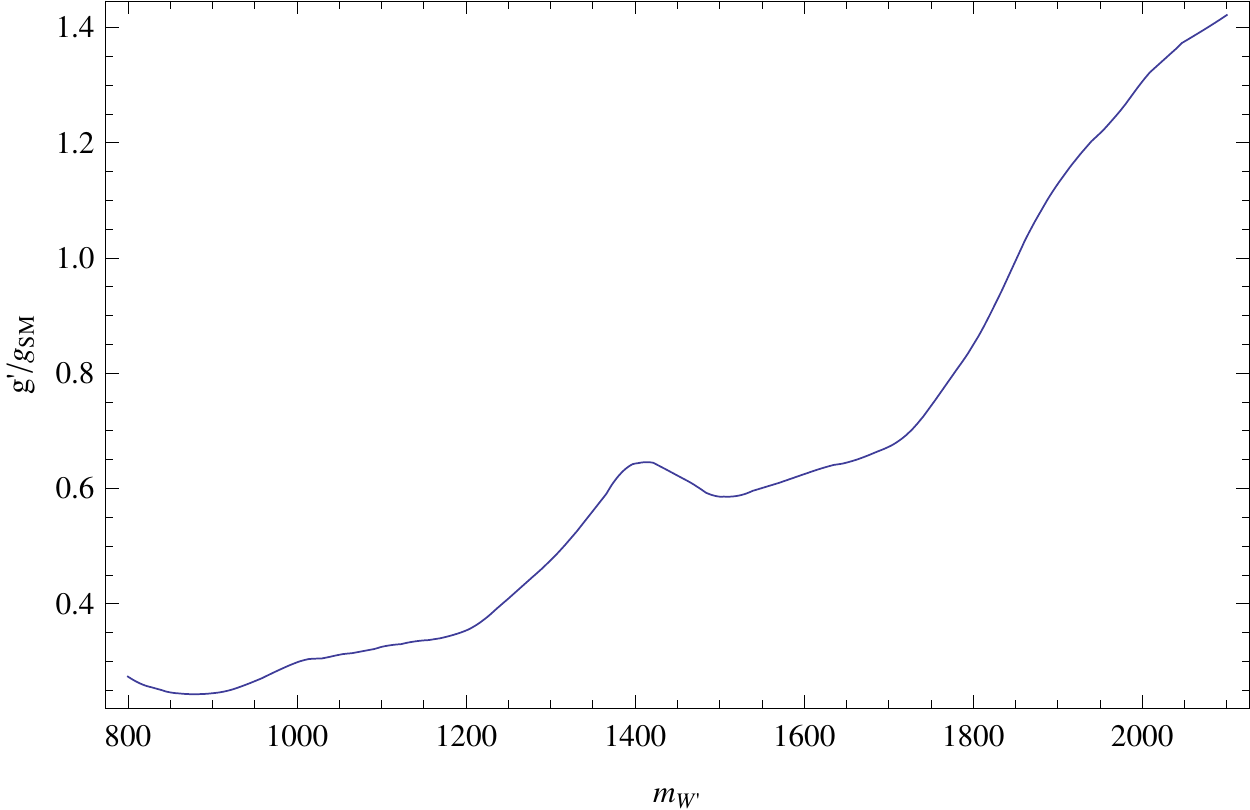}
\end{center}
\caption{Model independent bound on the relative $W^\prime$ coupling
  $g^\prime/g_{SM}$ vs.  $m_{W^\prime}$ at 95\% C.L. from the 5.0~fb$^{-1}$
  CMS data~\cite{CMSDATA}.}
\label{fig::Wpbound}
\end{figure}

In this section we utilize a constraint on $W^\prime$ boson production
and decay to an $s$-channel single-top-quark final state \cite{CMSDATA} to
place limits on the three distinct UED models discussed above.
Combining the cross section limit of Ref.\ \cite{CMSDATA} with the
predicted signal for a $W^\prime$ boson with standard model-like
couplings, we construct the bound shown in Fig.~\ref{fig::Wpbound}. We
find a model independent constraint~\cite{Sullivan:2003xy} on the
magnitude of $M_{W^{(2)}}$, and its coupling $g^\prime$ to zero-mode quarks.

\subsection{Bounds on MUED from $W^\prime$ searches}

Using Eq.~(\ref{g002mued}) we see that the couplings of $W^{(2)}$ to
gauge bosons is dependent on $M_{W^{(2)}}$ and $\Lambda$. Using
$\mu=2/R$ (mass of $m_{W^{(2)}}$) as a renormalization scale and
$g^2_3=4 \pi \alpha_s$ with $\alpha_s=0.12$, the relative coupling $\frac{g_{002}}{g_{000}}$ as a function of the
dimensionless cutoff $\Lambda R$ is given by \beq
\frac{g_{002}}{g_{000}}=-.065 \times \ln(\Lambda R/4)\,.
\label{gpogMUED}
\eeq 
With Eq.~(\ref{gpogMUED}), we can translate the bounds from $W^\prime$
searches displayed in Fig.~\ref{fig::Wpbound} into constraints on the
$\Lambda R$ vs.  $1/R$ MUED parameter space. However, due to the
logarithmic dependence on the compactification scale, only a very weak bound of  $\Lambda R \agt 100$ is obtained for the MUED model.\footnote{Taking the running of the strong coupling into account and evaluating the bound with $\alpha_s(\mu)$ leads to an even weaker constraint.}
This bound on $\Lambda R$ is weaker by an order of magnitude than bounds from existing searches \cite{NDAref,unitrefs,vacref}.

\subsection{Bounds on nUED from $W^\prime$ searches}

\begin{figure}
\begin{center}
\begin{tabular}{cc}
\includegraphics[width=.45\textwidth]{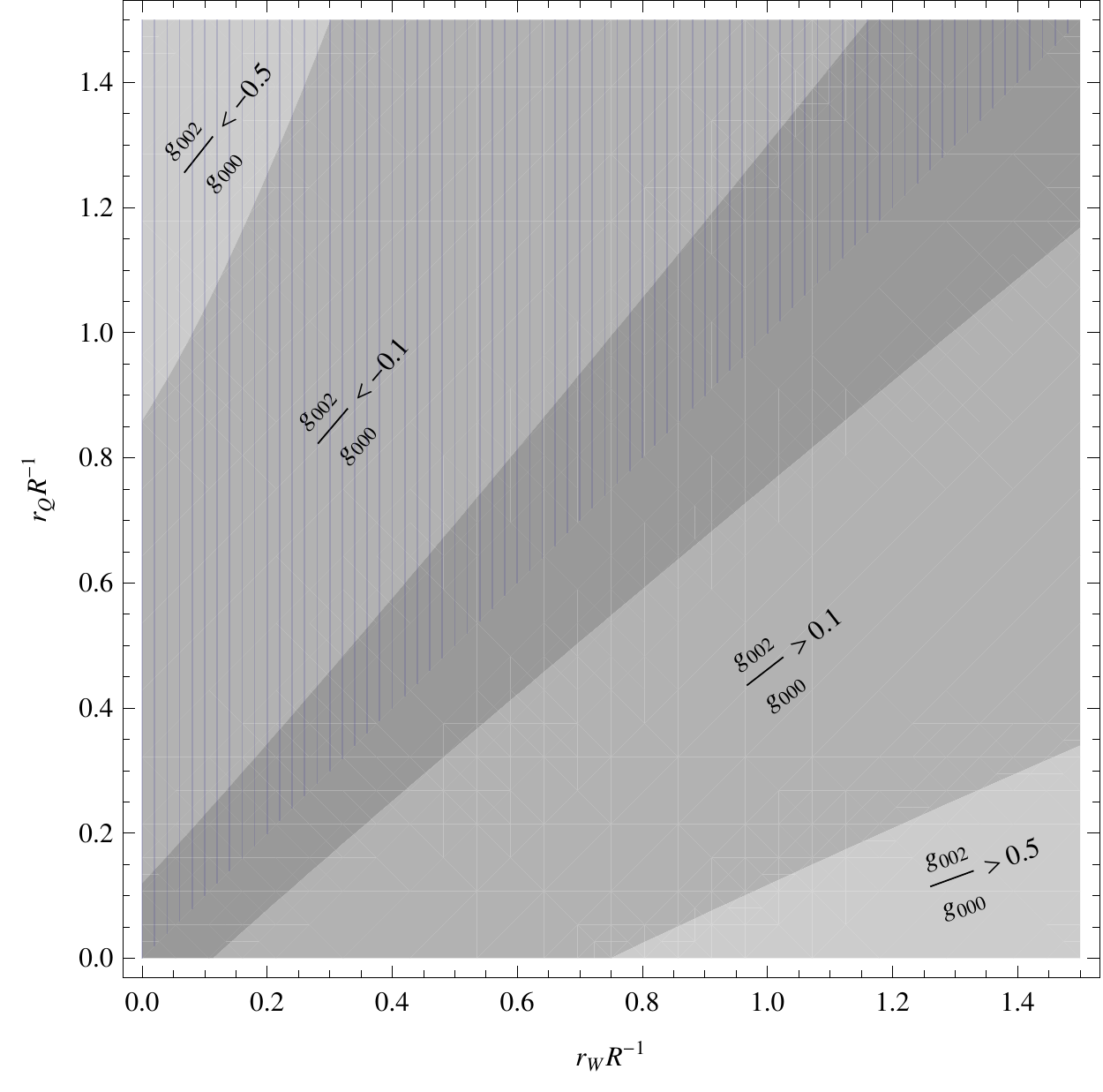}
\end{tabular}
\end{center}
\caption{Variation of the relative gauge coupling $g_{002}/g_{000}$ to quarks in
the $r_W R^{-1} - r_QR^{-1}$ plane.
\label{fig::nUEDgrel}}
\end{figure}

As can be seen from Eq.~(\ref{geffnUED}) and the determination of
$k_2$ in Eq.~(\ref{k2cond}), the ratio $g_{002}/g_{000}$ can be
expressed in terms of the dimensionless quantities $r_W/R$ and
$r_Q/R$. In Fig.~\ref{fig::nUEDgrel} we show the value of the relative coupling
$g_{002}/g_{000}$ in the $r_{W}R^{-1}$--$r_Q R^{-1}$ plane. As stated in
Sec.~\ref{sec:nUED}, when $r_W = r_Q$, KK-number violating terms vanish. 
$r_W>r_Q$ leads to positive $g_{002}$, which can even become larger than the standard model coupling. For 
$r_W<r_Q$,  $g_{002}$ is negative. When considering a common electroweak boundary parameter $r_H=r_B=r_W$, this parameter region is disfavored because the first fermion KK modes (here: the $Q^{(1)}$) are lighter than the usually considered dark matter candidate $B^{(1)}$. 
We shade this disfavored parameter region in
Fig.~\ref{fig::nUEDgrel}. If the electroweak boundary parameters are
not chosen equal, or if additional dark matter fields are included in
an extension of nUED, this region of parameter space can be opened up.

\begin{figure}
\begin{center}
\begin{tabular}{cc}
\includegraphics[width=.45\textwidth]{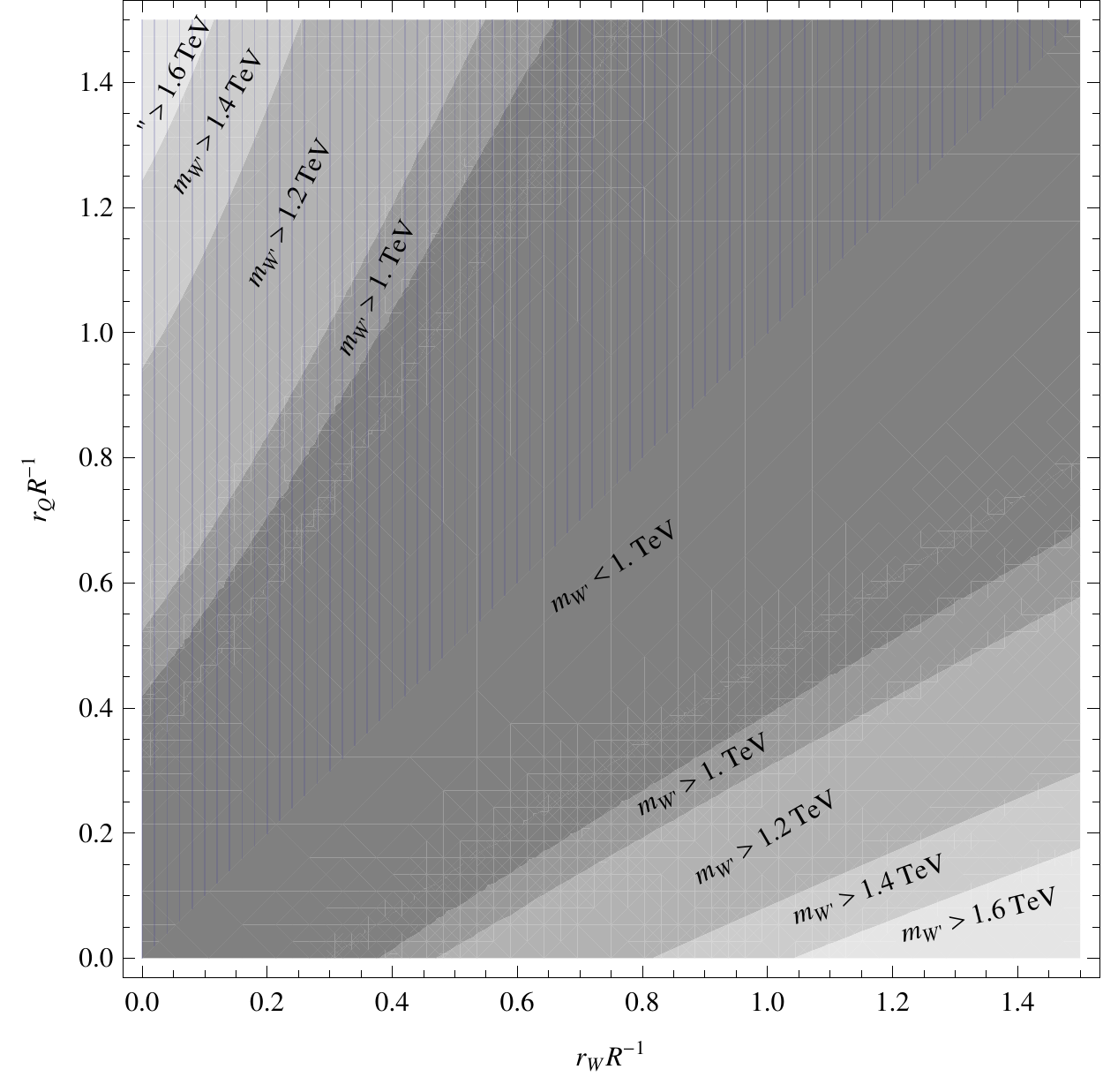}\\
\includegraphics[width=.45\textwidth]{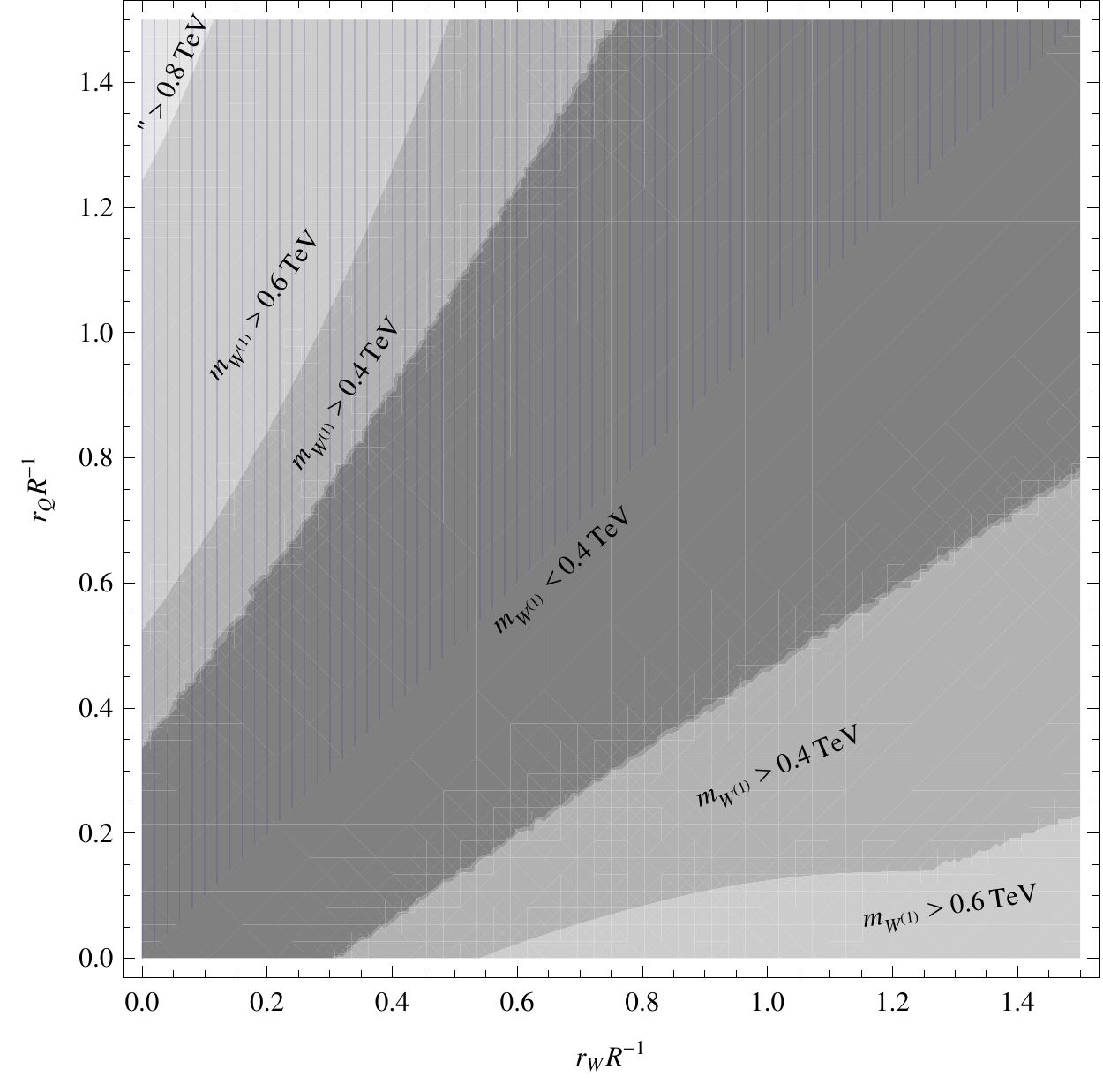}
\end{tabular}
\end{center}
\caption{Constraints $m_{W^{(2)}}$ (upper) and $m_{W^{(1)}}$ (lower) due to the CMS 
limit in Fig.~\ref{fig::Wpbound}.}
\label{fig::nUEDspec}
\end{figure}

Using Fig.~\ref{fig::nUEDgrel}, the $W^\prime$ limit in Fig~\ref{fig::Wpbound} can be translated into a constraint on the mass of the second KK excitation of the $W$ gauge boson $m_{W^{(2)}}\equiv m_{W^\prime}$. In the upper panel of Fig.~\ref{fig::nUEDspec},
we plot the limit on $m_{W^\prime}$ in
the $r_{W}R^{-1}$--$r_Q R^{-1}$ plane, where we have assumed a 100\% branching
ratio of $W^\prime$s to quarks.\footnote{Assuming a branching fraction to quarks similar to that of the standard model $\sim 75$\% does not significantly modify the contours, and smaller branching fractions to quarks are strong limited by precision electroweak constraints \protect\cite{FlackeP}.}

Similar to Fig.~\ref{fig::nUEDgrel}, the dark shaded region is
disfavored because the LKP would be the KK mode of a standard model
fermion.  Constraints are weak in the suppressed coupling region $r_W
\approx r_Q$, but become strong when the boundary parameters differ.

With the lower bound on $m_{W^{(2)}}$ in the upper panel of Fig.~\ref{fig::nUEDspec} and the nUED mass quantization conditions Eq.~(\ref{k2cond}) and Eq.~(\ref{k1cond}), a lower bound on the mass of each $W^{(n)}$ KK mode can be obtained. Of particular interest for LHC phenomenology is the first KK mode $W^{(1)}$.  If a common electroweak boundary parameter is assumed, its mass coincides with the mass of the $B^{(1)}$ LKP, up to a relative correction of the order $1-(m_W/m_{W^{(1)}})^2$, and is therefore relevant for dark matter bounds. In the lower panel of Fig.~\ref{fig::nUEDspec}, we translate the constraint on $m_{W^\prime} = m_{W^{(2)}}$ into a constraint on $m_{W^{(1)}}$. 

The constraints on the parameter space presented in Fig.~\ref{fig::nUEDspec} imply bounds on the allowed mass splitting between the first KK mode of the $SU(2)$ gauge boson $W^{(1)}$ and the $SU(2)$ charged quarks $Q^{(1)}$. For example, for a mass $m_{W^{(1)}}=600$~GeV, and a gauge BLKT of $r_WR^{-1}=1.0$, the minimally allowed value of $r_QR^{-1}$ can be read off from the lower panel of Fig.~\ref{fig::nUEDspec} to be $r_QR^{-1}\geq 0.13$. Using Eqs.~(\ref{Wnmass}) and (\ref{k1cond}),  the value of $R^{-1}$ is given by $R^{-1}=930$~GeV, which via Eq.~(\ref{k1cond}) yields $m_{Q^{-1}}\leq 860$~GeV, so that the relative mass splitting for these values of  $m_{W^{(1)}}$ and $r_WR^{-1}$ is given by $(m_{Q^{(1)}}-m_{W^{(1)}})/m_{W^{(1)}}\leq 45\%$.

An absolute bound on the mass splitting for a fixed $m_{W^{(1)}}$ mass, independent of the value of $r_WR^{-1}$ cannot be established in the nUED model, which can be seen as follows: In the limit $r_W R^{-1}\rightarrow \infty$, the relative mass splitting $(m_{W^{(2)}}-m_{W^{(1)}})/m_{W^{(1)}}\rightarrow \infty$ such that in this limit, $m_{W^{(1)}}$ can be kept constant while the $W^{(2)}$ mode decouples from the model. As the constraints discussed here arise from $W^{(2)}$ mode exchange, no bounds on $r_QR^{-1}$ are obtained in this limit.

\subsection{Bounds on split UED from $W^\prime$ searches}

The limits on $W^\prime$ masses and couplings, due to the search in the
single-top-quark channel, are especially important for the split-UED model
because it puts constraints on the quark bulk mass $\mu_Q$. The
original motivation for sUED is to raise the quark KK-mode
masses while allowing for light KK-leptons. Such a split spectrum was proposed
in Ref.~\cite{sUEDref} in order to explain the positron excess
observed by the PAMELA experiment while suppressing the anti-proton
rates.  Hence, in such models we expect large values of $\mu_Q$ and
small values of $\mu_L$.

\begin{figure}
\begin{center}
\includegraphics[width=.6\textwidth]{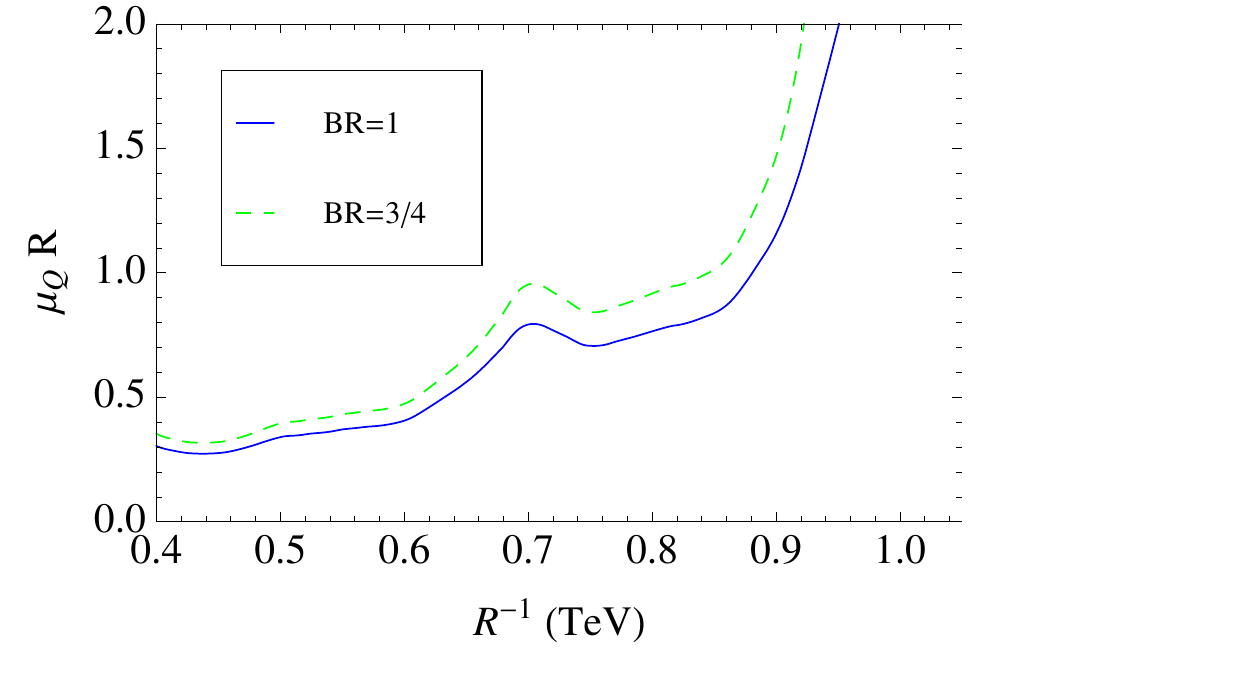}
\end{center}
\caption{ $\mu_Q R$ vs.\ $R^{-1}$ split-UED parameter space, where the contour 
  lines correspond to different branching ratios into quarks and leptons 
for the $W^\prime$ constraint shown in Fig.\ \protect\ref{fig::Wpbound}.}
\label{fig::sUED1}
\end{figure}

Using bounds from the $W^\prime$ search in the single-top-quark
channel shown in Fig.~\ref{fig::Wpbound} leads to constraints on the
$\mu_Q R$ vs.\ $R^{-1}$  split-UED parameter space shown in
Fig.~\ref{fig::sUED1}.
Depending on the magnitude of lepton bulk mass term $\mu_L$, the
branching ratio of the $W^\prime$ can vary, which is illustrated by
the different contour lines in Fig.~\ref{fig::sUED1}. The blue (dark
grey) contour is a scenario in which the $W^\prime$ decays only into
quarks, and the green (light grey) dashed 
contour is a scenario in which the
$W^\prime$ has branching ratios of 75\% to quarks and 25\% to leptons, similar to those of the standard model $W$ gauge boson.

As described in Ref.~\cite{Huang:2012kz}, constraints on
the four Fermi contact operator interactions and searches in
dileptons and dijets put constraints on the split-UED parameter space. The
dijet limit depends on the mass of the KK-gluon, which is not necessarily 
proportional to the the mass of the Kaluza-Klein partners of the
electroweak sector. Both the dilepton and the four Fermi contact
operator limits depend on the product of the couplings of the KK
partners of the electroweak sector to quarks and leptons.  Therefore
in the limit of small $\mu_Q$ or $\mu_L$ the dilepton and four Fermi
contact operator limits are weak. The $W^\prime$ search limit shown in
Fig.~\ref{fig::Wpbound} allows us to disentangle these effects and
puts orthogonal constraints on the $\mu_L$ versus $\mu_Q$ parameter
space.
 
\begin{figure}
\begin{center}
\includegraphics[width=.33\textwidth]{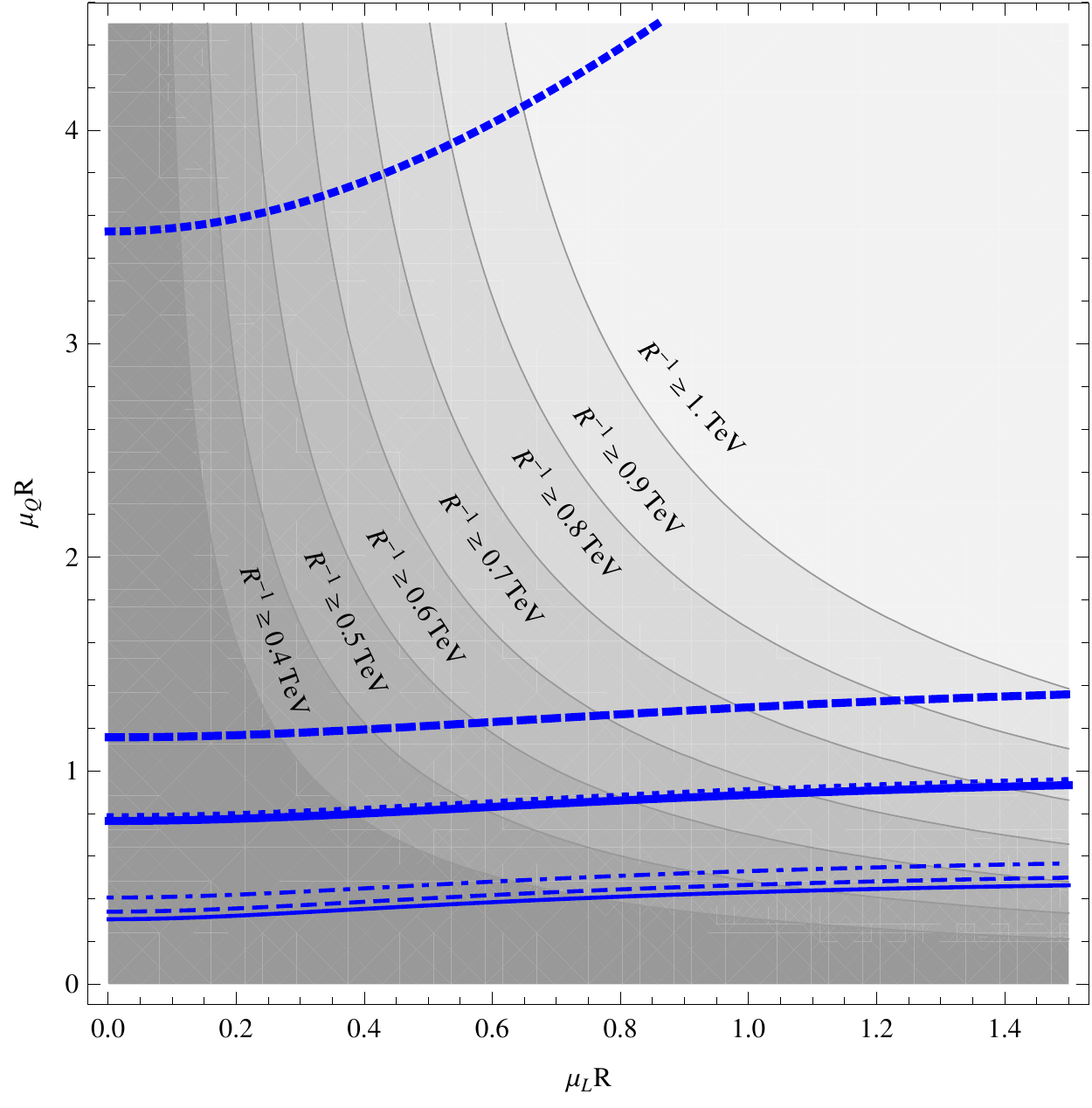}
\vspace{-5cm}
\includegraphics[width=.12\textwidth]{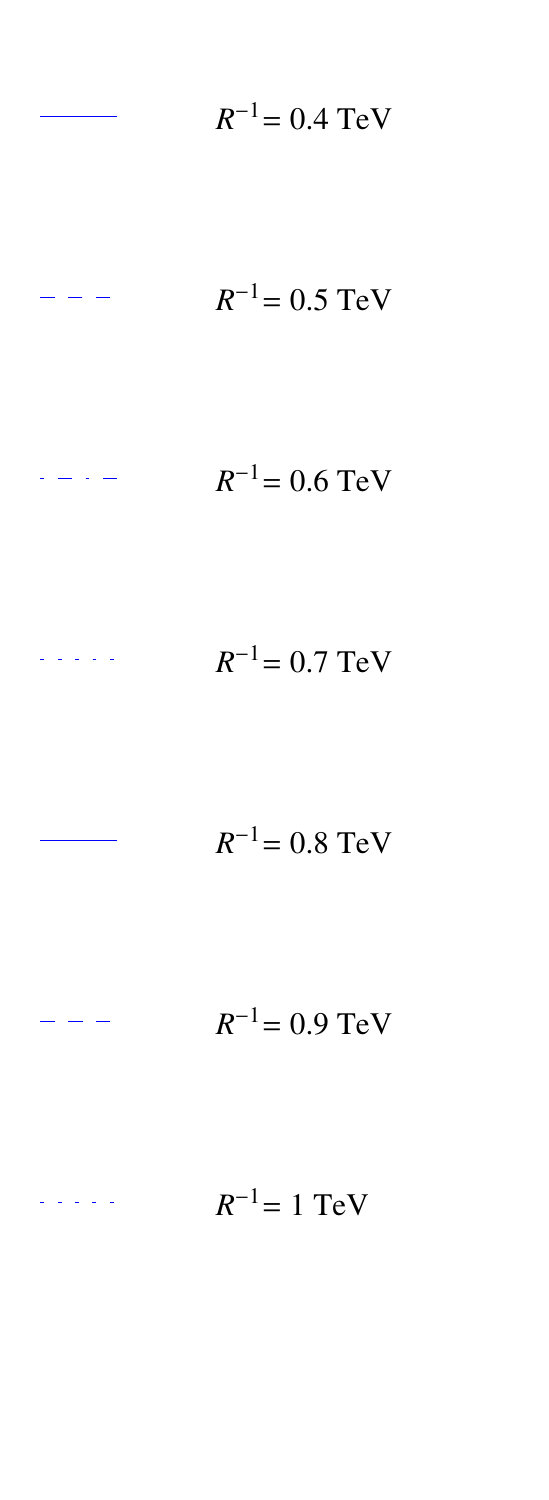}
\end{center}
\vspace{5.cm}
\caption{Limits on sUED parameter space due to the combination of the
four Fermi contact interactions constraint and 
the $W^\prime$ constraint displayed in Fig. \ref{fig::Wpbound}.}
\label{fig::sUEDcont}
\end{figure}

To illustrate the power of $W^\prime$ search limit shown in
Fig.~\ref{fig::Wpbound}, we combine it with the $eedd$ four Fermi
contact operator interaction limits of Ref.~\cite{Huang:2012kz} in
Fig.~\ref{fig::sUEDcont}. The grey contours correspond to the limits
on the $\mu_L R$--$\mu_Q R$ plane due to the $eedd$ four
Fermi contact interaction limit, while the horizontal lines are the
limits due to the $W^\prime$ prime search in the single-top-quark
channel. The slight weakening of the $W^\prime$ search limit for large
$\mu_L$ is due to the increasing branching ratio into leptons. The
$W^\prime$ search limit for $R^{-1} = 0.7$~TeV and $R^{-1} =
0.8$~TeV are comparable because of the slightly weaker constraint at
$1.4$~TeV in Ref.~\cite{CMSDATA}.

Just as we show for nUED, the bounds of Fig.~\ref{fig::Wpbound} can be
translated into bounds on the relative mass splitting
$(m_{Q^{(1)}}-m_{W^{(1)}})/m_{W^{(1)}}$.  We consider
$m_{W^{(1)}}=800$~GeV as an example.  For $\mu_L R=0$,
Fig.~\ref{fig::Wpbound} gives an upper bound of $\mu_Q
R=0.76$.\footnote{Choosing the $\mu_L R$ maximally allowed by the
  four-Fermi interactions leads to a slightly weaker constraint of
  $\mu_Q R=0.92$ for $m_{W^{(1)}}=800$~GeV. However, the case of
  $\mu_L R=0$, $\mu_Q R> 0$ is of particular interest for sUED dark
  matter searches, because in this limit, the dark matter annihilation
  rate into positrons is maximized while the production of
  anti-protons is maximally suppressed.} Using Eq.~(\ref{sUEDmW2}) and
Eq.~(\ref{sUEDmQ1}), we obtain a value of $m_{Q^{(1)}}\leq 1.0$~TeV and
hence a maximally allowed relative mass splitting of
$(m_{Q^{(1)}}-m_{W^{(1)}})/m_{W^{(1)}}\leq 25\%$.

\section{Conclusions}
\label{sec:concl}

In this paper we have shown that the $W^\prime$ limit from single-top-quark production leads to strong constraints on split-UED and 
the non-minimal UED models. For sUED, the $W^\prime$ limit puts a strong upper
bound on $\mu_Q$, the bulk mass parameter of the $SU(2)$ charged quarks. The upper bound on 
$\mu_Q$ implies an upper bound on the mass splitting between the $SU(2)$ charged  
KK-quarks and the $W$ KK excitations. This constraint is especially relevant, as the initial motivation 
for the sUED model required a large splitting between the KK-quarks and the LKP (whose mass scale is close to the $W^{(1)}$ mass) in order to suppress the production of anti-protons from dark matter annihilation at late times. 

In the nUED model, the coupling of the zero-mode quarks to the $W^{(2)}$ is 
induced by a splitting between the boundary localized terms. Hence the 
$W^\prime$ limit leads to constraints on the difference of $r_W R^{-1}$ and $r_Q R^{-1}$, which --- via Eq.~(\ref{k1cond}) --- again implies a bound on the mass splitting between $Q^{(1)}$ and $W^{(1)}$.

We emphasize that the $pp\rightarrow W^\prime \rightarrow tb$ channel
is particularly well suited to constrain the parameter space because
it only depends on the bulk quark mass parameter $\mu_Q$ in sUED and
the BLKT parameters of the $SU(2)$ charged quarks $r_Q$ and the
$SU(2)$ gauge bosons $r_W$ in nUED. Other mass terms or BLKTs have a
minor effect through altered branching ratios. This allows rather
robust bounds on the mass splitting between $SU(2)$ charged KK quarks
and KK $W$ modes to be obtained.  These bounds are robust because
production, as well as decay, of the $W^\prime$ are controlled by the
same coupling. Other search channels, like $Z^\prime$,
$\gamma^\prime$, $W^\prime$ in leptonic channels, or searches for
colored resonances depend on products of (linear combinations of)
different couplings. Allowing for generic bulk masses or boundary
terms therefore makes it more difficult to translate such searches
into particular mass splittings in the KK spectrum.

\begin{acknowledgments}
  A.M.\ and Z.S.\ are supported by the U.S.\ Department of Energy
  under Contract Nos.\ DE-FG02-96ER40969 and DE-SC0008347, 
  respectively. T.F.\ is supported in part by the Federal Ministry of Education and Research (BMBF) under contract number 05H09WWE.
\end{acknowledgments}

\end{document}